\documentclass[aps,prl,revtex4-1,twocolumn,superscriptaddress,showpacs,letter,longbibliography]{revtex4-1}

\usepackage{amssymb}
\usepackage{amsmath}
\usepackage{graphicx}% Include figure files
\usepackage{dcolumn}% Align table columns on decimal point
\usepackage{bm,textcomp}% bold math
\usepackage{braket}
\usepackage{hyperref}

\begin{document}

\author{Tobias Schaetz}
\affiliation{Albert-Ludwigs-Universit\"{a}t Freiburg, Physikalisches Institut, Hermann-Herder-Stra{\ss}e 3, 79104 Freiburg, Germany}
%\homepage[]{Your web page}
%\altaffiliation{}
%Collaboration name if desired (requires use of superscriptaddress
%option in \documentclass). \noaffiliation is required (may also be
%used with the \author command).
%\collaboration can be followed by \email, \homepage, \thanks as well.
%\collaboration{}
%\noaffiliation
\date{\today}
\title{Trapping Ions and Atoms Optically}
\pacs{37.10.Ty,03.67.Lx,34.50.Cx}
%\keywords{}

\begin{abstract}
Isolating neutral and charged particles from the environment is
essential in precision experiments. For decades, this has been
achieved by trapping ions with radio-frequency (rf) fields and
neutral particles with optical fields. Recently, trapping of ions by
interaction with light has been demonstrated. This might permit
combining the advantages of optical trapping and ions.
For example, by superimposing optical traps to investigate ensembles
of ions and atoms in absence of any radio-frequency fields, as well
as to benefit from the versatile and scalable trapping geometries
featured by optical lattices.
In particular, ions provide individual addressability, electronic
and motional degrees of freedom that can be coherently controlled
and detected via high fidelity, state-dependent operations. Their
long-range Coulomb interaction is significantly larger compared to
those of neutral atoms and molecules.
This qualifies to study ultra-cold interaction and chemistry of
trapped ions and atoms, as well as to provide a novel platform for
higher-dimensional experimental quantum simulations.
The aim of this topical review is to present the current state of
the art and to discuss current challenges and the prospects of the
emerging field.
\end{abstract}

\maketitle
%%%%%%%%%%%%%%%%%%%
\section{introduction}
Trapping charged atoms in rf-fields features deep trapping
potentials of the order of 1\,eV$\approx$\,$k_B\times 10^4$\,K and
related long lifetimes, as well as Coulomb interaction of long
range. Over decades, trapped ions and advanced tools for their
control are propelling many fields of
research\,\cite{Paul1953,Paul1990}. For example, progress in the
realm of quantum information processing results in unique control of
motional and electronic states of individual
ions\,\cite{Leibfried2003,Wineland2013}, located in one dimensional
Coulomb crystals, that is, linear strings of up to 20\,ions.
Operational fidelities close to unity for state preparation, phonon
mediated interactions and detection allow for reaching or setting
the state of the art for few particles, in quantum
metrology\,\cite{Rosenband2008,Chou2010,Huntemann2016}, quantum
computation\,\cite{Kielpinski2002,Wineland2011,Debnath2016,Monz2016},
as well as in analogue and digital quantum
simulation\,\cite{Friedenauer2008,Islam2011,Martinez2016,Blatt2012,Cirac2012}.\\
\begin{figure}[hhh]
\includegraphics[width = 0.35 \textwidth]{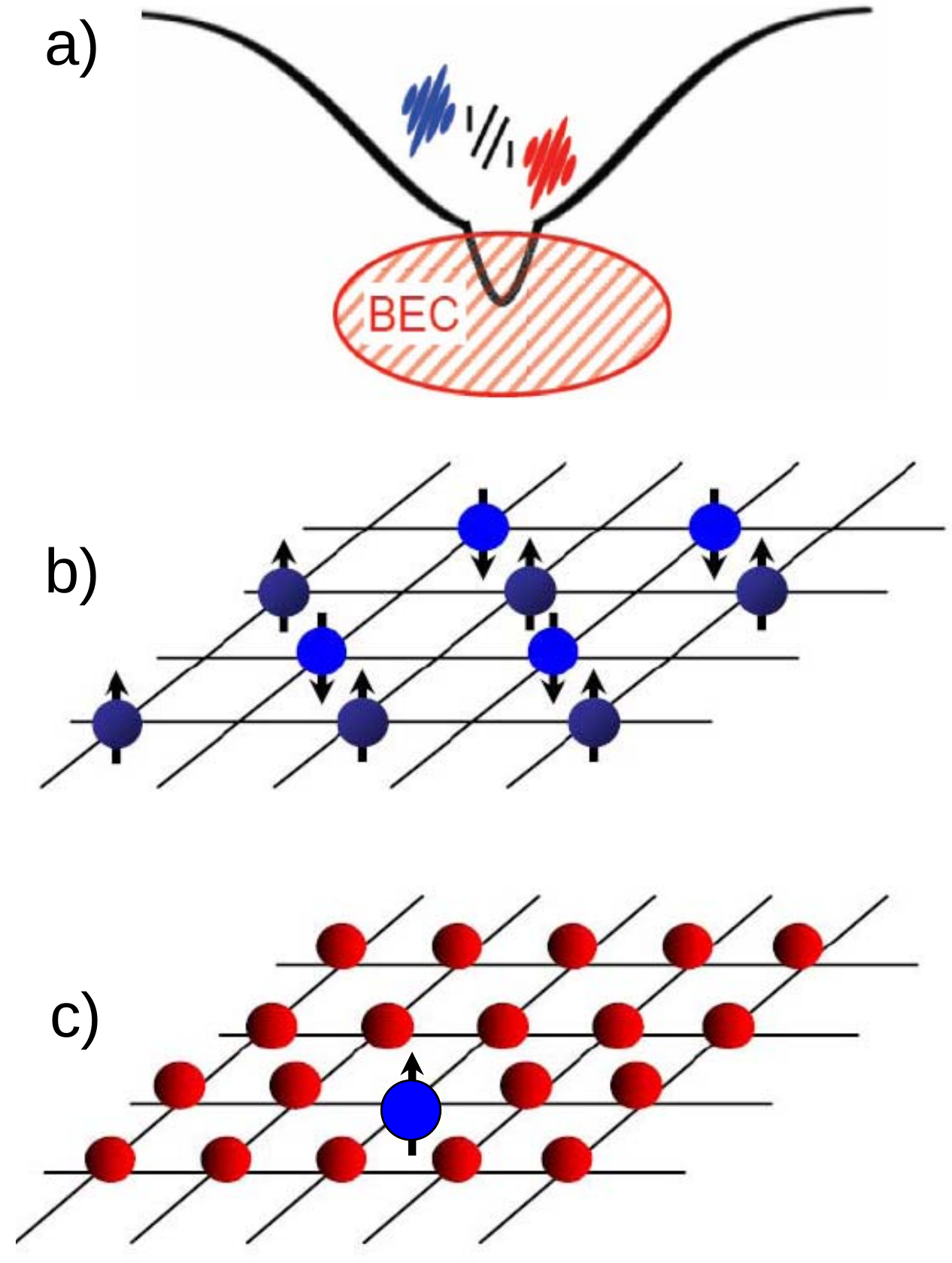}
 \caption{ (color online)
\textbf{Examples of applications for optically trapped ions:}
\textbf{a)} An ion immersed into an ultra-cold bath of atoms (BEC).
In the presence of rf-fields, the kinetic energy of the ion remains several
orders of magnitude above the temperature of the atomic bath.
Confining atoms and ions in a common optical potential
provided by a (bi-chromatic) dipole trap is predicted to feature
sufficient sympathetic cooling and, due to the absence of
rf-micro-motion, to lower the collisional energy down to the regime,
where quantum effects are predicted to dominate. 1D optical lattices
are envisioned to enrich this approach, (i) acting as conveyor
belts, on the single ion and single atom level or, (ii), providing
sympathetic cooling prospects by isolating generic (molecular) ions
from direct contact with its reactive coolant.
\textbf{(b,c)} New options for analogue quantum simulations (AQS) based on ions
(blue) and atoms (red) in optical potentials (black lines as a guide
to the eye) allowing for studies of many-body effects\,\cite{Schneider2012b}.
(b) Ions populate an optical lattice on well separated sites. The
Coulomb force inhibits filling of all lattice sites, but
still provides a large strength of dipolar, long-ranging
interaction, mediating effective spin-spin interaction.
(c) An ion and atoms populate a common, however, species and electronic state-dependent trap array
provided by an optical lattice.
This enables for
example, sharing the charge via tunneling electrons.
  \label{fig1}}
\end{figure}
However, trapping via rf-fields is accompanied by inevitable side
effects such as rf-driven motion. This so called micro-motion is
superimposed on the secular motion of the ion within its
time-averaged trapping potential and is undesirable or detrimental
for several
applications\,\cite{Cetina2012,Tomza2015,Chou2010,Hauke2012}. In
addition, scaling Coulomb crystals in size and dimension while
preserving the level of individual control remains a demanding
task\,\cite{Kielpinski2002,Mielenz2016,Wineland2013}.\\
For neutral particles, optical fields are a well established
technique for confining in potential wells of the order of $ k_B
\times 10^{-3}$\,K\,\cite{Grimm2000}, featuring versatile trapping
geometries scalable to large
two and three dimensional lattices\,\cite{Bloch2005}.\\
Combining advantages of ions and optical traps has been achieved by
a number of different approaches.\\
Spatially overlapping optical fields for trapping neutrals and
rf-fields for trapping charged atoms to study the atom-ion
interaction has lead to seminal experiments in so called hybrid
traps\,\cite{Smith2014}. In the temperature regime of tens of
millikelvin, reactive collisions involving molecular ions have been
studied\,\cite{Hall2012,Willitsch2015}.
To reach even lower temperatures, rf-trapped ions were immersed into
clouds of ultra-cold
atoms\,\cite{Grier2009,Schmid2010,Zipkes2010,Meir2016}, such as
Bose-Einstein condensates (see figure\,\ref{fig1}a). Hybrid traps
permit to prepare and investigate electronic- and motional states of
the ion using state of the art techniques\,\cite{Meir2016}. However,
in the context of atom-ion interaction, it has been shown that the
minimal energy of an ion embedded in a bath of ultra cold atoms
remains independent of the temperature of the bath and is set to the
level of 100\,$\mu$K. It is limited by the inevitable
rf-micro-motion induced during the atom-ion interaction, pumping
energy into the system (see figure\,\ref{fig2}) - even in the ideal
case where no stray electric fields are present\,\cite{Cetina2012,Nguyen2012,Tomza2015,Krych2015}.\\
Another approach is to provide the trapping of ions, replacing the
confinement along the axis of a linear rf-trap by a one dimensional
optical lattice. It permits locally shaping trapping potentials
while keeping the rf-confinement for the radial degrees of
freedom\,\cite{Katori1997,Linnet2012,Linnet2014,Karpa2013}. This
approach opened new perspectives, e.g., for studies of classical and
quantum phase transitions in the context of
friction\,\cite{Bylinskii2015,Bylinskii2016,Gangloff2015}.\\
The approach emphasized in this review focusses on trapping ions in
optical fields, omitting any rf-field. This has been achieved for
the first time in a close detuned optical dipole
trap\,\cite{Schneider2010} in the ultra-violet (UV), followed by a
one dimensional optical lattice\,\cite{Enderlein2012}. Recently,
trapping of ions in a far detuned optical trap (VIS) has been
reported\,\cite{Huber2014}.
Further improvements resulted in
%an increase of
lifetimes of a Doppler cooled ion
%by two
%orders of magnitude via active optical re-pumping in the (FORT at
%532nm) and
reaching several seconds, exploiting enhanced control and a further
de-tuned laser
(NIR)\,\cite{Lambrecht2016}.\\
Long lifetimes of ions in dipole traps, accompanied by long coherent
times and low heating rates are beneficial to reach the regime of
ultra cold interaction in dilute atom-ion ensembles and key for
experimental quantum simulations in optical lattices with many ions,
or ions and atoms, as
anticipated in figure\,\ref{fig1}b,c.\\
In the following, we aim to discuss challenges and prospects of the
field on the basis of two showcases:\\
(1) in the field of ultra cold chemistry: Embedding optically
trapped ions into quantum degenerate gases is predicted to allow for
reaching temperatures, 4-5 orders of magnitude below the current
state of the art (see figure\,\ref{fig2} and
references\,\cite{Cetina2012,Nguyen2012,Tomza2015}. This should
permit entering the regime where quantum effects are predicted to
dominate: (i) in many-body physics, including the potential
formation and dynamics of mesoscopic clusters of atoms of a
Bose-Einstein-Condensate, binding to the ``impurity
ion''\,\cite{Cote2002}, as well as (ii) the subsequent two-particle
s-wave collisions, the ultimate limit in ultra-cold chemistry.\\
(2) in the field of analogue quantum simulations (AQS): Optically
trapping of ions in lattices might short cut the current lack of
scalability in size and dimension of the trapped ion systems. In
addition, new classes of quantum simulations will become accessible,
by combining optically trapped ions and atoms in
common optical lattices (see figure\,\ref{fig1} and references\,\cite{Schneider2012b,Blatt2012}).\\
%
%
%We experimentally select and analyze the state dependent (trapping
%repelling) optical potentials. In addition, we measure the
%current upper bound for the heating rate in our apparatus caused by
%laser, rf and ambient fields and reveal that it might already permit
%entering the ultra cold regime via sympathetic cooling of atom-ions
%species of a variety of mass ratios.
%
\section{Methods}
\begin{figure}[t]
\includegraphics[width = 0.48 \textwidth]{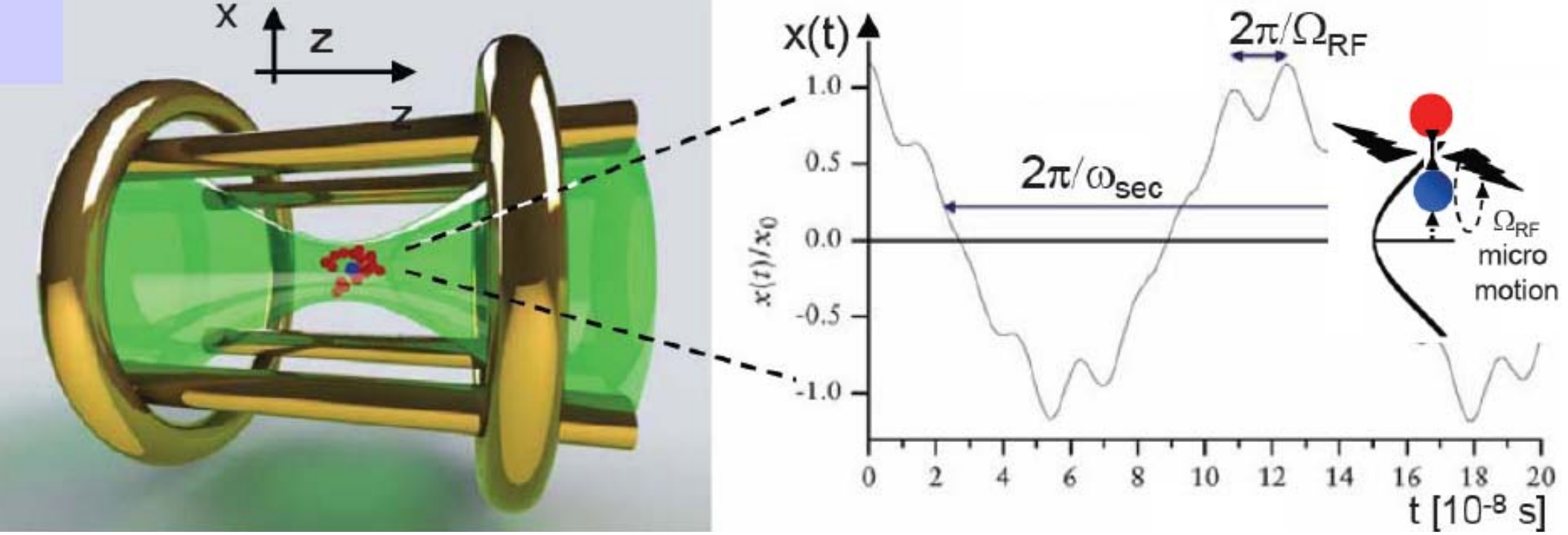}
 \caption{ (color online)
\textbf{rf-micro-motion and the related, fundamental heating of an atom-ion ensemble:}
left: Spatially overlapping the center of a linear rf-trap for ions
(gold: four quadrupole rods providing the radial-rf, two ring
electrodes the axial-dc confinement) with an optical dipole trap
(green) for atoms. right: The rf-field forces the ion to quiver at
the rf-drive of frequency $\Omega_{RF}$ when it is displaced from the rf-node (and/or
oscillates with its secular frequency $\omega_{sec}$ in the
time-averaged pseudo potential).
inset: Even if the ion is taken to behave as a classical object, at
T=0 and assuming ideally compensated stray-electric fields, the
intrinsic interaction between the ion and an approaching atom leads
again to an inevitable displacement from the rf-node. The time-dependent disturbance via the rf field
sets the fundamental lower limit to the minimal achievable mutual kinetic
energy.
  \label{fig2}}
\end{figure}

We want to summarize fundamental questions on the general
limitations and challenges of optical trapping of ions, as well as
the methodology required.\\
While preparation and detection remains provided by a conventional,
linear rf-trap, optical trapping is performed in the presence of
electric stray and control fields, but in the absence of any
rf-field. The experiments so far followed a common
protocol\,\cite{Schneider2010}, further detailed in sections below.
An atom out of a thermal beam is resonantly photo-ionized and
trapped in a linear Paul trap. The confinement is provided by
rf-fields for the radial degrees of freedom and by dc-fields along
the axis (see figure\,\ref{fig2}). After trapping and cooling of the
ion, (1) the Doppler cooling laser is switched off, (2), the dipole
trap laser is switched on, and (3), the RF potential is ramped down
to zero. The dipole trap is kept on for the optical trapping
duration. Afterwards, the sequence is reversed for the transfer back
into the conventional Paul trap. If the CCD camera monitors the
fluorescence light of the ion during Doppler cooling again, it
reveals with near unity efficiency that the trapping attempt has
been successful. Note, during the transfer between the different
traps, all fields have to be considered simultaneously.
\subsection{coupling of external fields to the charge monopole\\
 and the optically induced dipole}
We discuss the consequences of the charge monopole on the trapping
characteristics within the optical trap that relies on the optically
induced dipole. This includes the ion's dynamics during the
intermittent exposure to the rf-dc-optical potentials while getting
transferred between traps and the potential ``onset'' of differences
in optically trapping ions compared to optically trapping of
atoms\,\cite{Grimm2000}.
To the best of our knowledge, all relevant aspects have been
considered and elaborated on\,\cite{Cormick2011} and we identified
the main, technically remaining limitations, stray-electric fields.
All topics are briefly summarized in the following - our current
solutions are listed, while their details can be found in the
specified references.
For the case of stray electric fields we present our current
working-solution in more detail in the following subsection.
\\
(1) The effect of the light field directly interacting with the
charge causes a residual, optically-driven
micro-motion\,\cite{Cormick2011}. Calculations show that this causes
a negligible impact since it remains further detuned by seven orders
of magnitude compared to the rf. For an atom-ion ensemble, it is
predicted to permit reaching
the pico-Kelvin regime.\\
(2) The impact of constant or slowly drifting stray electric fields
on the charge disturbs trapping conditions\,\cite{Haerter2013a} and
can shift the ion out of the shallow optical potential, independent
on the temperature of the ion\,\cite{Schneider2012}. To permit
decreasing laser intensities to reduce the related off-resonant
scattering and heating effects,
stray-electric fields have to become further diminished (see below).\\
(3) The impact of the curvature of the dc-fields reduces the overall
trapping potential inevitably by de-focussing\,\cite{Schneider2012}
following to Laplace's equation\,\cite{Paul1990}. These dc-fields
are required for providing dc-confinement in some degrees of
freedom, controlling curvatures (trapping frequencies) and
orientation of the principal axes\,\cite{Kalis2016}, while allowing
for additional diagnostics and
control.\\
(4) The dynamics caused by switching optical and rf -fields, e.g.,
to control the transfer of the ion between traps, can affect the
stability of the individual traps and cause consequences for the ion
exposed to their intermittent
overlap\,\cite{Enderlein2012,Schneider2012}. On the one hand, the
rf-potentials should be switched sufficiently fast to cross higher
order resonances within the stability regime of the idealized
quadrupole potential\,\cite{Alheit1995} without relevant energy
transfer. On the other hand, they must be switched sufficiently slow
in comparison to the inverse of the relevant trapping frequencies to
enable an adiabatic transfer. In addition, optical lattices might
feature trapping frequencies of the order of half of the frequency
of the rf-drive, causing direct or parametric heating. Intermittent
storage within a dipole trap and subsequently switching to the
lattice in the absence of the rf turned out to
be viable\,\cite{Enderlein2012}.\\
(5) The trapping conditions degrade if a frequent reloading of both
species of choice is required. Higher background pressure reduces
lifetimes and compromises stable conditions, either by lethal
collisions, enhanced by higher ion-atom cross sections compared to
their neutral counter parts\,\cite{Wineland1998} or
non-deterministic near-crossing trajectories of residual ions
tramping in the rf-trap. Cleaning the trapping volume completely,
e.g.\,, by switching off the rf-field while saving a single ion by
an intermittent dipole trap turned out to be helpful. Ablative
loading via a pulsed laser system can help to reduce this
problem\,\cite{Hendricks2007} accompanied by efficient
photo-ionization\,\cite{Leschhorn2012}. Further improvement is
anticipated following the concept of conveyor belts for single,
sympathetically cooled molecular
ions\,\cite{Kahra2012,Leschhorn2012b} up to crystalline, three
dimensional beams of ions\,\cite{Schramm2002}, as established in
this context for atoms\,\cite{Zipkes2010,Haerter2014,Meir2016},
passing differential pumping sections. Cleaner conditions could get
be further assisted by a cryogenic environment. These measures will
also help to reduce the deposition of patch potentials, beneficial
to decrease the impact of effects treated in (2) and
below.\\
(6) The effect of technical noise out of the environment can be
mitigated by efficient filtering, ideally down to the level where
fluctuating patches on the surfaces of the electrodes cause
anomalous heating\,\cite{Wineland2013}. This is currently not a
limiting factor for experiments, since these effects drop with
$d^{-4}$, where $d$ represents the distance to the electrode.
However, for even reduced laser intensities and related trapping
frequencies, these effects might again influence the ion when aiming
for lowest temperatures. Recently, methods for reducing anomalous
heating by more than two orders of magnitude with either surface
treatments~\cite{Allcock2011,Hite2012,daniilidis2014} or cold
electrode
surfaces~\cite{deslauriers_scaling_2006,labaziewicz_suppression_2008,labaziewicz_temperature_2008}
have been reported. The fundamental limitation will be set by
Johnson noise, related to the thermal motion of the electrons within
the metal, decreasing with decreasing temperature of the
electrodes\cite{Hite2012}.
The heating due to beam pointing and frequency/intensity jitter of
the lasers involved should remain comparable to their effects on
neutral atoms, however, the ensemble will remain prone to a mutual
jitter of lasers within a bi-chromatic trap.
%
%(7) The different combinations out of many possible atom and ion
%choices, addressable by our generic approach (leading to Rb-Ba$^+$
%and a promising alternative, Li-Ba$^+$).
%
%
\subsection{compensation of stray fields\\
 exploiting the ion as the sensor}
The state-of-the-art techniques for compensating stray-electric
fields had to be improved by two orders of magnitude, to
(1) set the stage for the compensation of excess rf-micro-motion in
hybrid traps containing Li-atoms\,\cite{Cetina2012} and to
(2) further improve the optical trapping conditions in the future,
for example, to permit optically trapping at substantially lower
laser intensities\,\cite{Huber2014}.\\
We have to consider that trapped ions will be displaced by residual
stray-fields. Optical trapping is possible for a maximal
displacement of half of the beam waist, where Gaussian beams yield
the steepest gradient and their maximal counteracting force,
respectively (see figure\,\ref{fig3}a).
Note that optical forces acting on the ion can remain comparable to
the trapping forces of the minimal rf-confinement obtained just
before the rf-field is switched off, even though the total depth of
the trapping potential shrinks by 6-7 orders of magnitude.
Our method of choice is based on minimizing the displacement of the
ion while operating the rf-trap\,\cite{Huber2014}, even though the
rf-drive is irrelevant after switching the rf-trap off. The approach
exploits the ion itself as a sensitive detector to identify
displacements due to residual fields by switching the depth of the
rf-pseudo-potential between extremal amplitudes. Exploiting the
tightly focussed dipole beam itself, yet strongly attenuated, still
induces an differential AC-Stark shift. It alters the fluorescence
rate of the ion dependent on its position and relates it to the two
stray-field dependent positions for two different rf-confinements.
In addition, the method allows optimizing the overlap of the centers
of the rf-trap and the optical trap(s), minimizing the heating
during an imperfect adiabatic transfer between the traps. Modulating
the rf-confinement and exploiting lock-in techniques are promising
to further enhance the sensitivity.
\subsection{protocol of optical trapping:\\
rf-trap for established preparation/cleaning and detection only}
The procedure for optical trapping a $^{24}$Mg$^+$ and
$^{138}$Ba$^+$ ion within UV
(280\,nm)\,\cite{Schneider2010,Enderlein2012}, VIS
(532\,nm)\,\cite{Huber2014} and NIR (1064\,nm)\,\cite{Lambrecht2016}
dipole traps is identical. The direction of the k-vector has been
varied, enclosing between $45^\circ$ and $0^\circ$ with the z-axis
of the linear trap (see figure\,\ref{fig2}), realized in three
different apparati\,\cite{Schmitz2009,Leschhorn2012b}.
To load an ion into a dipole trap and to finally detect its optical
trapping consists of the following steps:\\
(1) Photo ionizing an atom out of a thermal atomic beam, trapping,
Doppler-cooling it in a conventional rf-trap and appropriately
compensating stray-fields.\\
(2) Providing the dipole trap by a Gaussian laser beam focussed onto
the ion and switching off the rf-drive of the Paul trap. From that
time on, the ion is confined by the dipole trap in the directions
perpendicular to the k-vector of the beam. The conventional dipole
trap is assisted by dc-confinement along the axis of the linear
trap, while the 1D optical lattices provided by retro-reflecting the
beam confines an ion all optically. Loading might have to be
assisted by
an intermittent step (see issues of parametric heating treated above).\\
(3) Switching on the rf-drive again and verifying the presence of
the ion via its fluorescence during Doppler-cooling.\\
AC-Stark shifts and the related optical potentials and forces depend
on the wavelength of the dipole laser and the dedicated electronic
state\,\cite{Grimm2000}.
For $^{24}$Mg$^+$, a red detuned UV dipole laser provides trapping
in the S$_{1/2}$ ground-state.
The VIS trap on $^{138}$Ba$^+$ provides trapping in the S$_{1/2}$,
but anti-trapping in the  D$_{3/2,5/2}$ states. The latter can
either become occupied by optical pumping or reached by off-resonant
scattering, leading to a loss of the optically trapped ion.
Recently, a re-pumping scheme was realized, allowing for
de-populating the  D$_{3/2,5/2}$ states during optical trapping,
prolonging the lifetime of the ion by an order of
magnitude\,\cite{Lambrecht2016}.
The NIR trap on $^{138}$Ba$^+$ provides trapping in the S$_{1/2}$
and in the D$_{3/2,5/2}$ states, the latter still confined, but at
trap depths
reduced by $\sim$\,75\%\,\cite{Lambrecht2016}.\\
We determine  (i) the lifetime, (ii) the temperature, (iii) heating
rates and (iv) scattering rates, by measuring the optical trapping
probability in dependence on different parameters. To access (i), we
vary the duration of the trapping attempt, for (ii) we alter the
optical trap depth and for (iii) we follow the protocol of the
temperature measurement, however, adding a delay between preparation
and optical trapping. During this delay trapping is still secured by
rf-fields, however, it incorporates the specific operation to
investigate its impact. To derive (iv), we study
the influence of re-pumping lasers.\\
In addition: There is the huge variety of control, diagnostic and
detection schemes which are available from work on small numbers of
ions in rf-traps\,\cite{Wineland2013}.
Methods, such as (sympathetic) cooling to the motional ground
state\,\cite{Barrett2003}, individual
addressing\,\cite{Schaetz2004}, state detection with close to unit
efficiency down to single photon sensitivity\,\cite{Clos2014},
coherent state operations\,\cite{Rosenband2008,Schneider2012b}, as
well as vibrational spectroscopy with femto-second resolution on
single molecular ions\,\cite{Kahra2012} are at hand already, still
awaiting their exploitation.
\section{Results}
\subsection{trapping the ions optically}
In the year 2009, optical trapping of a $^{24}$Mg$^+$ ion, in
absence of any rf-fields, was achieved (optical power $\sim$0.2\,W
at a waist of\,$\sim$6.5\,$\mu$m) - first in a single-beam dipole
trap, superimposed by a static electric
potential\,\cite{Schneider2010}. Subsequently, all optical trapping
of the ion in an optical lattice (1D-standing wave) has been
demonstrated\,\cite{Enderlein2012} for similar parameters.
However, the lifetime of the ion remained limited to milliseconds.
This was due to off-resonant photon scattering out of the trapping
laser field and the related photon recoil. At a recoil energy of
2E$_{recoil}\approx 10\mu$K and a scattering rate of 750\,ms$^{-1}$,
the ion is heated out of the
comparatively shallow trapping potential of a depth of the order of 40\,mK.\\
One-dimensionally pinning of ions(s) in rf-hybrid traps was reported
in the year 2010 for $^{40}$Ca$^+$\,\cite{Linnet2012} and in 2012
for $^{174}$Yb$^+$\,\cite{Karpa2013}. Radially, the confinement of
the ion(s) along one axis was still achieved by an active linear
rf-trap, however, overlapped with the axis of a standing wave within
an optical cavity.
Side-band cooling close to the motional ground state has been
achieved within these novel hybrid traps \,\cite{Karpa2013}.\\
In 2014, a purely optical confinement of a $^{138}$Ba$^+$ ion within
a far-detuned dipole trap was demonstrated (optical power
of\,$\sim$10\,W at a waist of\,$\sim$4\,$\mu$m). This allowed to
reduce recoil heating by four orders of magnitude, however, the
lifetime remained limited to milliseconds. This is still due to the
off-resonant scattering rate: indeed, the latter is strongly reduced
by three orders of magnitude, but remains responsible for optical
pumping into meta-stable electronic states of Ba$^+$, featuring
different AC-Stark shifts and repelling forces induced
by the VIS dipole laser.\\
Still, the lifetime is sufficient to immerse Ba$^+$ ions into an
ensemble of ultra-cold atoms within a common dipole trap and to
investigate sympathetic cooling already.
To permit efficiently enhanced life- and coherence times in dipole
traps and optical lattices, following the receipts of trapping
neutral atoms, the detuning as well as the laser power has been
increased\,\cite{Grimm2000}.
In 2016, trapping of a Ba$^+$ ion within an IR dipole trap was
achieved\,\cite{Lambrecht2016}. This is the key prerequisite to
permit confinement of ultra-cold atoms and ion(s) in a common
(bi-chromatic)
optical trap (see figure\,\ref{fig3}).\\
Considering the compensation of stray-electric fields, we are
currently reaching the level of 1\,mV/m at the position of the ion.
This level permits lowering the intensity of the lasers and is
sufficient to address and test prediction on the suitability of Li
to reach the quantum regime in hybrid traps\,\cite{Cetina2012}.
\section{Discussion}
The aim of this section is to discuss the significance of the
results reported above to help circumventing fundamental limitations
or substantial challenges in applications by combining the
advantages of optical trapping and ions.
First, we discuss requirements and prospects of optically trapping
in the regime of temperatures directly reached by Doppler-cooling.
Second, we dwell on the improvement of our approach in the context
of aiming at reaching the regime of ultra-low temperatures. The
related prospects for some applications, illustrated in
figure\,\ref{fig1}, will be
described in the next section.\\

To increase the lifetime, lower laser intensities and related
scattering rates are beneficial. Currently, deep optical traps are
still required to permit trapping ions at comparatively high initial
temperature.
A promising option to reduce the residual loss is to choose an ion
species with a reduced branching ratio into meta stable states or
even featuring a closed transition only. For some applications,
e.g.\,, those dependent on a larger amount of optically trapped
ions, it might be sufficient to increase the rate of re-pumping by
stroboscopically alternating laser exposure at a rate much larger
than the inverse of the secular frequencies.
%$ \gg \omega_{x,y,z}$, however,
Still, residual scattering compromises longer coherence times.
\subsection{Coulomb crystals in optical traps}
\begin{figure}[t]
\includegraphics[width = 0.48 \textwidth]{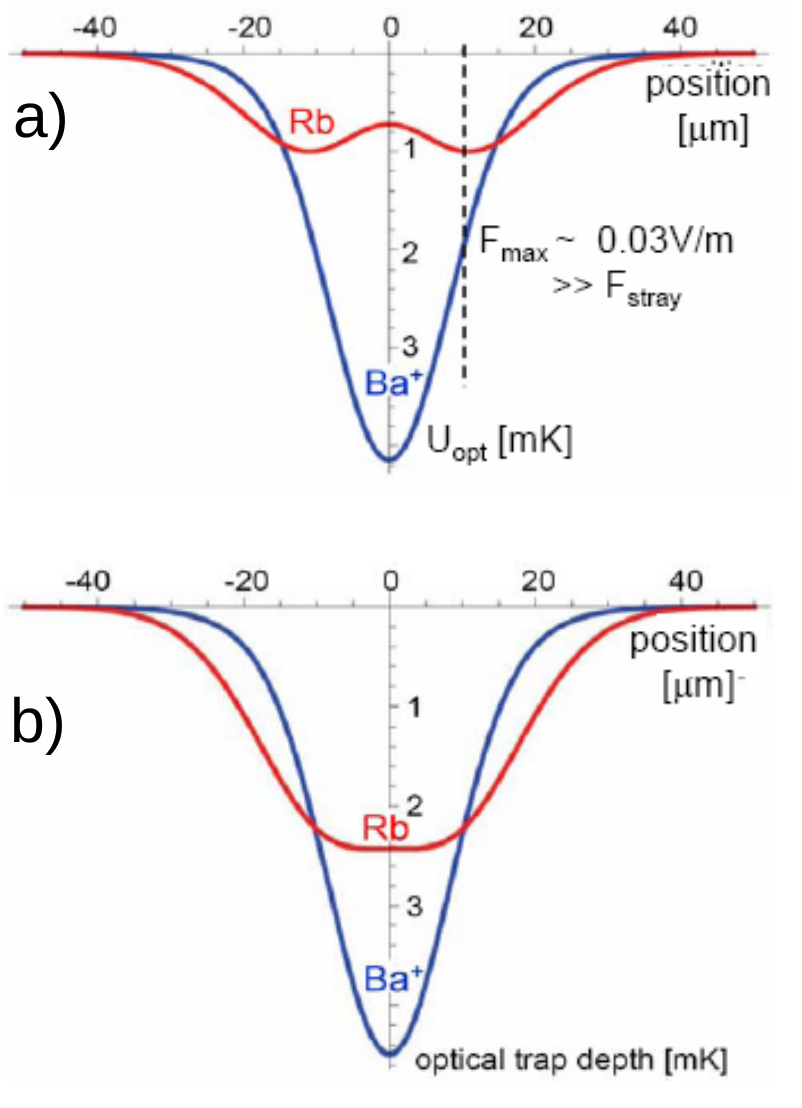}
 \caption{ (color online)
\textbf{Optical trapping potentials for $^{138}$Ba$^+$ (blue) and Rb (red) for different, available
laser powers within the bi-chromatic (1064\,nm and 532\,nm) dipole trap of given waists:}
\textbf{a)} The 532nm beam is responsible for trapping Ba$^+$ (493\,nm) and is leading to
a de-focusing effect in the bi-chromatic trap centre for Rb (780nm), mainly
stored by the 1064\,nm wavelength, spatially separating the two species.
\textbf{b)} Increasing the IR power permits compensating the de-focussing effect
and to spatially overlap atoms and the ion in the centre. Adjusting
the relative laser power will allow to tune the depth of the optical traps
individually, e.g., to enable and control the evaporative and sympathetic cooling rates.
Preparing sufficiently low temperatures by side-band and sympathetic cooling might again turn
common trapping within the NIR trap advantageous.
\label{fig3}}
\end{figure}
To illustrate the protocol for entering the regime of ultra-cold
atom-ion ensembles, we select the combination of a $^{138}$Ba$^+$
with two atomic species of extremely different masses, $^{87}$Rb and
$^{6,7}$Li, here focussing on Rb.
Long lifetimes are indispensable for scaling to a larger number of
ions to investigate multiple ions trapped in optical potential(s).
For example, for studying Coulomb crystals and structural phase
transitions to more dimensional structures at Doppler temperatures,
well established and exploited in
rf-traps\,\cite{Drewsen2015,Thompson2015}. However, more dimensional
Coulomb crystals in rf-traps experience an even enhanced impact of
the rf-field on the ions, since the ions are intrinsically displaced
from the rf-free axis. In this context, the state dependency of
optical trapping potentials and the available coherent control of
electronic states can be exploited to follow intriguing proposals on
structural phase
transitions\,\cite{Baltrusch2012,Baltrusch2013,Cormick2012,Shimshoni2011}
and related quantum effects. Optionally, embedding sympathetically
cooled (molecular) ions or even sympathetically cooled, positively
charged ions, sufficiently close, but spatially separated to ions of
negative charge will enrich the prospects.
%
%The purely optical approach circumvents the currently inevitable
%excess kinetic energy in hybrid traps, where ions are kept but also
%driven by rf-fields. It might permit to enter the temperature regime
%where quantum effects are predicted to dominate. The prospects for
%in many-body physics, including the potential formation and dynamics
%of mesoscopic clusters of atoms of a Bose-Einstein-Condensate,
%binding to the "impurity ion",
%
%In parallel, we explore further perspectives by adapting methodology
%of quantum optics to gain control and state-sensitive detection on
%the level of individual quanta within the merged ion-atom system.\\
%
%In this context Further development of novel and generic tools for
%quantum engineering can be expected to propel several other striving
%Fields of research, such as, experimental quantum simulations.
%
%%%%%%%%%%%%%%%%%%%
%%%%%%%%%%%%%%% Entering the Ultra-cold regime
%%%%%%%%%%%%%%%%%%%
%
%Relevance of the results in the context of
%
\subsection{Cooling atom-ion ensembles into the quantum regime}
To provide the required bath of ultra cold atoms, we follow
established procedures\,\cite{Arnold2011}.

(a) Realizing an all optical, bi-chromatic trap for $^{138}$Ba$^+$
ions and $^{87}$Rb atoms: Loading of Rb atoms into a MOT (typically
$2\times10^8$ atoms ($\sim 10^{11}cm^{-3}$ at 150\,$\mu$K -
50\,$\mu$K after an optical molasses - with a lifetime of $>$10\,s)
followed by an all-optical BEC ($4\times10^4$ atoms after 3\,s of
loading and evaporative cooling to 50\,nK temperatures).
It has been demonstrated that Ba$^+$ ions can get be trapped in VIS
and NIR dipole traps\,\cite{Lambrecht2016}. Combining the two
provides a bi-chromatic dipole trap that permits controlling the
confinement of the Rb atoms separately
%at 1064\,nm and 532\,nm
(see figure\,\ref{fig3}).
Specifically, it permits inducing species dependent (de)focussing
and, thus, additional control on the spatial overlap, the density of
the atomic ensemble in the vicinity of the ion. The example shown in
figure\,\ref{fig3}a depicts a deep confinement of earth alkali ions
in the centre of the bi-chromatic trap, while a dedicated choice of
beam parameters provides a Mexican hat like potential for the
alkaline atoms due to the VIS laser remaining detuned to the blue of
their relevant transitions. Simpler choices, such as, the
confinement of the atom-ion ensemble by the NIR laser alone would
provide deeper confinement for the atoms, however, it might also
compromise
sympathetic cooling during evaporation of atoms.\\
To reduce loss mechanisms while sympathetically cooling the ion
embedded in the cold atomic ensemble, e.g., due to three body
collisions\,\cite{Kruekow2016} (see also figure\,\ref{fig4}), might
require initially reduced atomic densities and related extended
periods of elastic atom-ion interaction. In this context it might
become essential that the measured upper bound of the heating rate
of an ion in the presence of trapping and ambient fields remains
sufficiently low, for example, in comparison to predicted
sympathetic cooling rates, as estimated in\,\cite{Krych2011}.
Reaching sufficiently low temperatures at a first stage will permit
the option to trap the ensembles in the NIR trap.
Fully exploiting the current level of stray field compensation (E\,$
\approx$\,0.001\,V/m) might permit substantially lower laser
intensities\,\cite{Huber2014}.\\
(b) Sympathetic cooling $^{138}$Ba$^+$ by $^{87}$Rb (BEC) down to
100\,nK:
The atomic system is predicted by theory to be suitable for
sympathetically cooling the ion-atom ensemble to the temperature of
the atomic bath on the time scale of
milliseconds\,\cite{Krych2011}.\\
Several methods are at hand to measure the rate of sympathetic
cooling and the final temperature, such as, deriving the energy
distribution of the ion by fitting optical trapping probabilities
assuming the energy cut-off model\,\cite{Schneider2012}, side-band
diagnosis\,\cite{Wineland2013,Meir2016} and analyzing the loss rate
of the atoms, as
demonstrated in the hybrid traps already\,\cite{Haerter2014}.\\
(c) Sympathetic cooling $^{138}$Ba$^+$ by $^6$Li and $^7$Li -
comparison of trapping via rf- versus optical fields:
It should become possible to elucidate the prospects of Li atoms as
(i) sympathetic coolant, advantageous since most efficient due to
its small mass,
(ii) collision partner, offering different isotopes of bosonic and
fermionic nature,
(iii) insensitive to charge transfer, since the relevant inelastic
reactions in the Li-Ba$^+$ system are expected to remain endothermic
(Li and Ba$^+$ forming the ground state), and
(iv) remaining the currently only atomic candidate species not
fundamentally excluded from exploring and exploiting the ultra cold
regime in hybrid traps.
In this context it is worth mentioning that it has recently been
argued that the first few collisions might also get probed in rf
traps for a wide range of species\,\cite{Meir2016}.
%%%%%%%%%%%%%%%%%%%
%%%%%%%%%%%%%%% Cold Interaction -
%%%%%%%%%%%%%%%%%%%
%
\section{conclusion}
In this section, we aim to further highlight the novelty of the
approach by concluding with plans for future work.
\subsection{Ultra Cold interaction - quantum many-body effects}
\begin{figure}[t]
\includegraphics[width = 0.48 \textwidth]{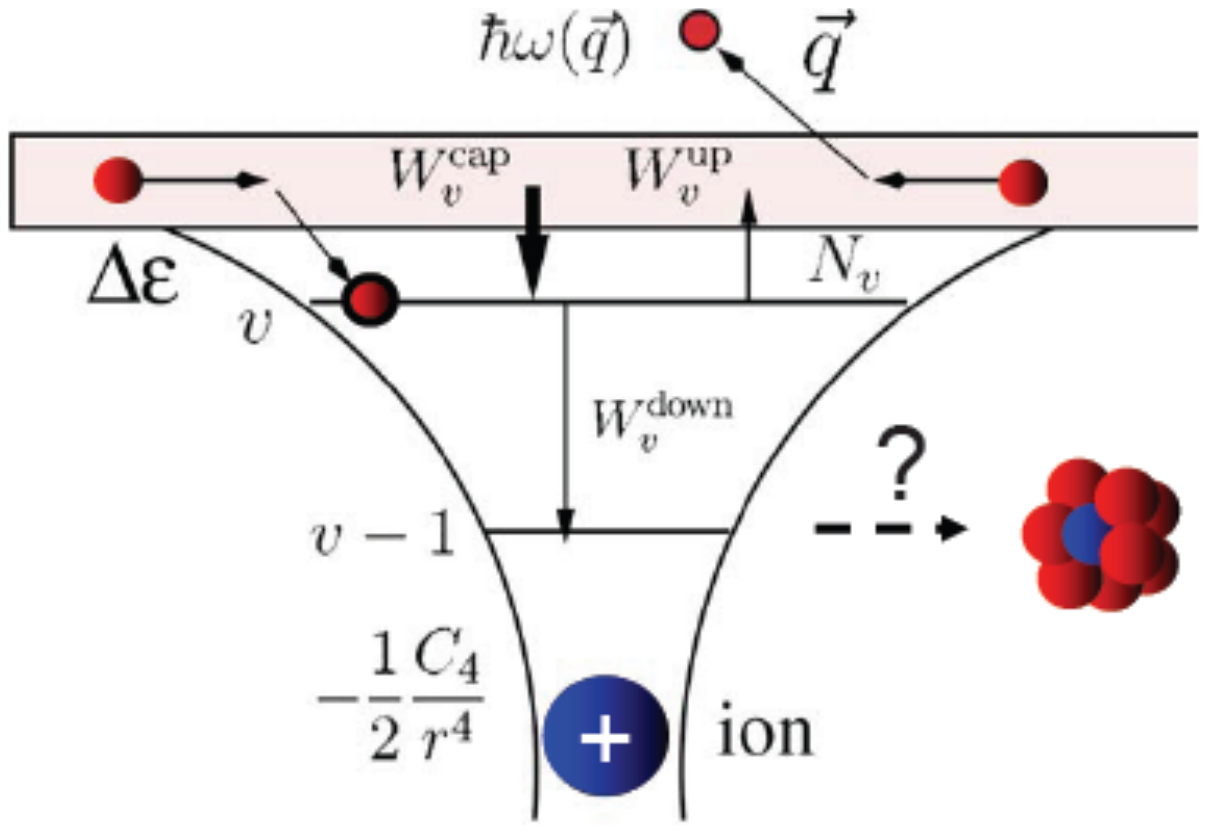}
 \caption{ (color online)
\textbf{Schematics of the predicted cluster formed by an ion binding
many atoms within a BEC:}
Diagram of atom capture by an ion where the spontaneous capture in level v is followed by phonon emission
at the corresponding rates $W_v^{cap,up,down}$
(curtesy of M.\,Lukin and R.\,Cote from\,\cite{Cote2002}). The process of formation might be accompanied
by additional losses due to the increased density in vicinity of the ion. Additional control is envisioned via
Feshbach resonances and two-photon stimulated Raman transitions.
  \label{fig4}}
\end{figure}
Optical trapping should enable exploring and directly observing
many-body processes in the atomic BEC-ion system within the
ultra-cold regime. For example, the test of the related predictions
on the rapid and efficient formation of a metastable and mesoscopic
molecular ion, binding of the order of hundreds of atoms of the BEC
within the common ion-atom $r^{-4}$-potential (see
figure\,\ref{fig4} and reference\,\cite{Cote2002}).
It has to be noted that the related atomic densities within these
objects at the predicted sub-micrometer diameter would raise to
extremely high values. That is, the dynamics might be strongly
affected by three-body collisions and related
losses\,\cite{Kruekow2016}, requiring a description beyond the mean
field theory the predictions are currently based on. Additional
mutual dipolar repulsion of the polarized atoms or even active
control via simultaneous atom-atom and predicted atom-ion Feshbach
resonances might be considered\,\cite{Idziaszek2011,Tomza2015}.
The system might allow revealing the characteristics and related
dynamics of the ion-atom system during cooling to the nano-Kelvin
regime, exploiting tools to control parameters and detect
observables of the many-body system directly.
Gaining insight into the underlying processes, as well as into the
prospects for exploring and exploiting the control might relate
these results to solid state/cluster physics; relevant, for example,
for processes in condensed matter systems and liquid helium where
so-called snowballs and electron-bubbles have been
observed\,\cite{Cole1974}. There are substantial advantages of the
atomic BEC-ion system. For example, the low atomic density should
permit comparatively clean experiments, while winning established
techniques from the field of quantum optics permit highly efficient
detection and control for the novel system. One of the key features
of an heterogeneous approach is the ability to distinguish the
Ba$^+$ from the atomic Rb-BEC, thus, eliminating resonant charge
transfer.
Potential tools are:\\
(a) measuring the secular frequency, i.e., effective mass of the
immersed ion: classical vibrational spectroscopy\,\cite{Kalis2016}
will allow a direct comparison of the secular frequency of the ion,
bare and immersed into ultra cold atoms. Any interaction, especially
the potential binding of atoms to the ion, will affect the
oscillation frequency of the ion in the optical trap. Thus,
measuring an altered frequency can be interpreted as observing a
dressed ion featuring an effective mass related to the amount of
bound atoms, e.g., via resonant excitation and the related enhanced
loss rate of atoms out of
the surrounding ensemble.\\
(b) measuring motional excitation on the quantum level:
phonon-spectroscopy via two-photon stimulated Raman transitions can
be exploited to (i) implement side-band cooling to the motional
ground state of Ba$^+$, e.g., for initial state preparation, and to
(ii) measure the population of the quantized motional states,
permitting direct measurements of the temperature and energy
distribution of the ion, respectively\,\cite{Meir2016}.
Additionally, the related motional control can be extended on the
predicted ion-atom cluster to (iii) gain external control on the
raising and lowering rates of the atomic population in the bound
states. For specific energies/detuning, and as the atoms surrounding
the ion drop deeper into the ``ionic well'', the remaining BEC atoms
would gain energy and be ejected from the trap, leading to quantized
atom loss rates for discrete laser detunings - potentially assisted
by a modulation of the trapping potential. Thus, implementing (iii)
could permit to spectroscopically resolve the binding energy,
chemical potential and energy difference between bound
states\,\cite{Cote2002}. Additionally, driving the transitions at
dedicated Rabi-rates might permit to reveal the timescales for the
formation of the cluster, as well as measuring and controlling the
internal state dynamics on their path from an originally metastable
(all bound atoms in the highest excited state) towards its deeper
bound states - and extract the dependence on relevant parameters of
the system, such as, its
finite size and temperature.\\
(c) detection of the interaction range of polarization via EIT:
recently, EIT has been shown suitable to image the blockaded Rydberg
sphere in cold atomic samples\,\cite{Amthor2010}, permitting a
non-destructive detection of Rydberg excitations. The interaction
between two Rydberg atoms is reported to become sufficiently strong
to shift the relevant energy levels out of resonance: illuminating
the atoms by a pair of EIT-lasers reveals a shift of levels,
breaking the former condition for transparency, causing the region
affected to become opaque again. A similar approach should be
applicable in the atom-ion systems: in the region surrounding the
ion, the atoms experience a mutual dipole-dipole interaction and the
polarization interaction due the ion that might shift the relevant
levels sufficiently to break the EIT condition, hence leading to a
way to image the location and size of the related cluster. Combined
with the method described in (a), the atomic density within the
ion-atom
ensemble  might become accessible.\\
(d) exploiting the sensitivity of the ``ion-impurity'' to external
fields: depending on the success of the methodology described above,
the control on the BEC might be extended itself, using the ion as an
inherent impurity, precisely addressable by external fields.
Applications might be (i) creating and controlling rotational
excitations (vortices) within the BEC, (ii) controlling the spatial
position of the BEC by electrostatic forces and altering the
trapping conditions, and (iii) immersing more than one ion into the
BEC causing a potential partial shielding of their charges and
mutually reduced
repulsion, e.g., to further extend (i) and (ii).\\
%%%%%%%%%%%%%%%%%%%
%%%%%%%%%%%%%%% Cold Chemist
%%%%%%%%%%%%%%%%%%%
%
\subsection{ Ultra Cold Chemistry - reactive collisions}
Dependent on the dynamics occurring on the way towards the
ultra-cold regime, two-particle s-wave collisions can be reached.
The ultimate limit in ultra-cold chemistry might become accessible
in fundamentally different ensembles. Examples are:
(1) Ba$^+$ ion and Rb: fundamentally out of reach for hybrid traps.
(2) $^{138}$Ba$^+$ and $^{6,7}$Li: a priori not excluded for hybrid
traps, assuming a sufficient level of stray field compensation.
\begin{figure}[t]
\includegraphics[width = 0.4 \textwidth]{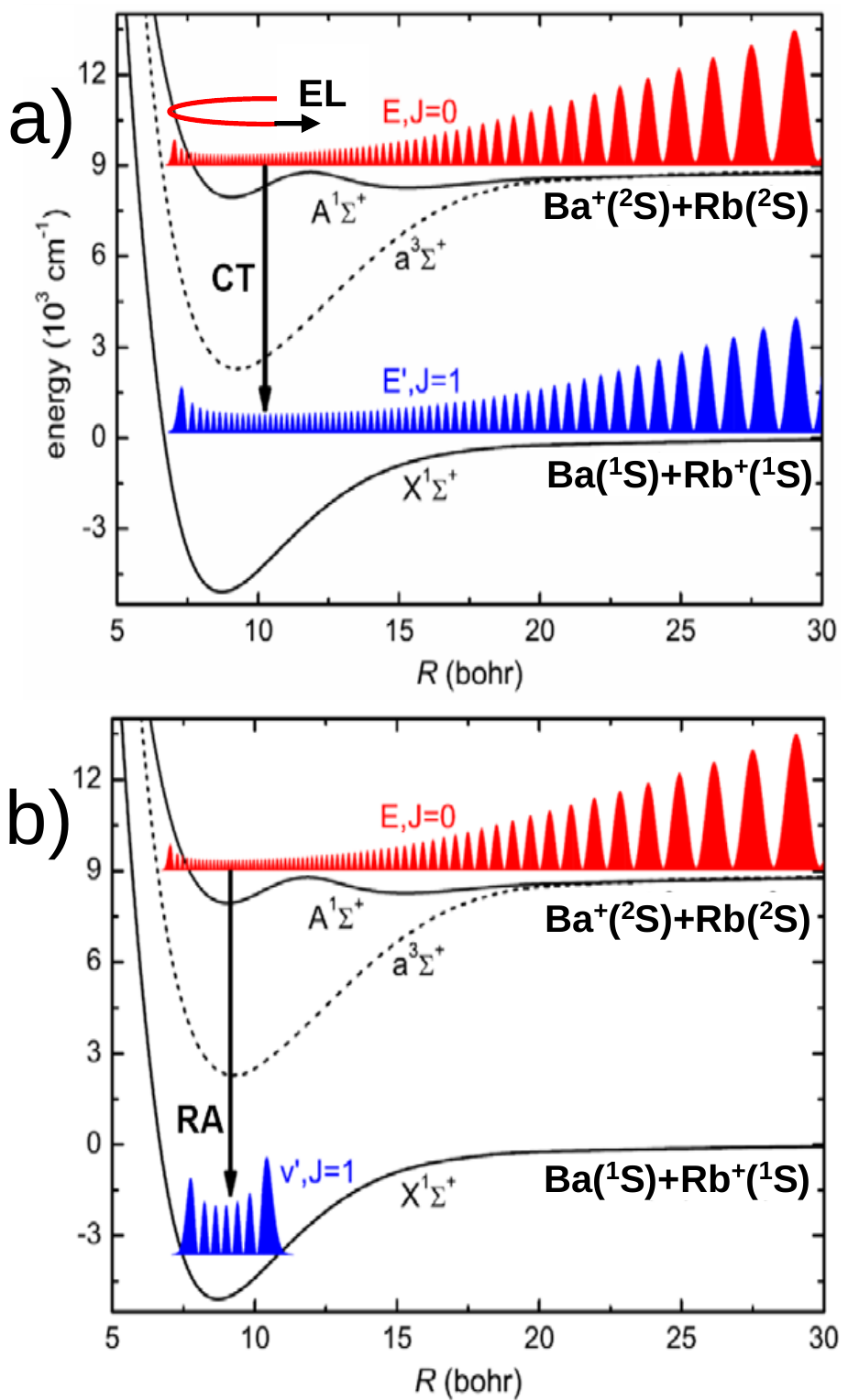}
 \caption{ (color online)
\textbf{Schematics illustrating the two competing, inelastic processes,
relevant for the formation of a molecular ion:}
\textbf{a)} charge transfer (CT) and \textbf{b)} radiative association (RA),
shown here into one out of several ro-vibrational states. To permit
investigating the formation of molecular ions, the cross section for
RA has to be dominant (data of ab-initio calculations: courtesy of
R.Moszynski) and confirmed by experiments at milli-Kelvin
temperatures. Optically trapping might permit measuring the
cross-sections and state-sensitively detect the branching ratios
into the product states while keeping the molecular product, e.g.,
within a novel form of an hybrid trap, an intermittent rf-multi-pole
trap providing deep confinement at a negligible rf-field and related
minuscule rf-micro-motion at its centre.
  \label{fig5}}
\end{figure}
%
%Thus far, this has only been reached by one neutral-neutral system
%(KRb+KRb), however, without direct detection of the product and its
%quantum states, where the signature of the reaction became the
%disappearance of the
%product.\\
%
%Implementing the tools developed for trapped ions for control and
%detection should permit exploring the prospects of generic ion-atom
%systems to gain insight into the quantum scattering properties.
%
%This field of research is currently still at its infancy,
%theoretically as well as experimentally, mainly because the
%"long-range" $(1/r^{-4})$ polarization interaction is sufficiently
%strong to cause a comparatively large "bare" scattering length and a
%high density of states near the threshold. Gaining knowledge based
%on the envisioned extended amount of observables should enable the
%exploration of more complex phenomena.\\
%
Here, we briefly discuss the Ba$^+$-Rb system (see
figure\,\ref{fig5}). The collision process between the Ba$^+$($^2$S)
ion and the Rb($^2$S) atom involves the elastic scattering (EL)
between the ion and the atom, and two types of inelastic events:
charge transfer (CT) and radiative association (RA). Elastic
collisions will lead to the initial energy transfer from the ion to
the cold atom(s), that is, sympathetic cooling of the ion and the
subsequent escape of the atom(s) from the trap. This process will
also yield ultra-cold ions in first place, if the cross section for
the elastic scattering was much larger than that for the inelastic
events (predicted to be fulfilled for Rb and
Ba$^+$\,\cite{Krych2011}). Molecular states of the (BaRb)$^+$
molecule, corresponding to the Ba$^+$($^2$S)+Rb($^2$S) dissociation
limit, are excited states of the molecular ion (the ground state
corresponds to the Ba($^1$S)+Rb$^+$($^1$S) asymptote, lying
approximately 8344 cm$^{-1}$ below the Ba$^+$($^2$S)+Rb($^2$S)
asymptote). The collision between the Ba$^+$($^2$S) ion and the
Rb($^2$S) atom can lead to the formation of the bound (BaRb)$^+$
molecular ion in its electronic ground state via RA. The continuum
wave function of the singlet and triplet manifold, corresponding to
the Ba$^+$($^2$S) + Rb($^2$S) dissociation limit, is connected to
(several) bound ro-vibrational levels of the ground electronic state
through the electric dipole operator. A different inelastic
mechanism that may occur in the collision between the Rb($^2$S) atom
and the Ba$^+$($^2$S) ion involves the CT of the electron from the
Rb($^2$S) to the latter. Finally, a free Ba($^1$S) atom and
Rb$^+$($^1$S) ion will arise. The continuum wave function of the
excited triplet and singlet manifold is again electric dipole
connected to the ground electronic state,
but this time to the ro-vibrational continuum.\\

Close collaboration with theory might allow mapping out the relevant
processes as a function of the experimental conditions (e.g.,
energy/temperature, electric/magnetic fields, preparation of
internal states of atoms/ions).
It is worth emphasizing that optical fields can influence the
formation and dissociation of molecules, respectively. Dependent on
the wavelength, they might assist the formation of molecules by
radiative association or transfer population into a dissociative
channels\,\cite{Hall2011,Hall2012}.
Experimental data is required for understanding and ultimately
controlling the behavior of mixed atom-ion systems in the quantum
regime.
%
%%%%%%%%%%%%%%%%%%%
%%%%%%%%%%%%%%% AQS
%%%%%%%%%%%%%%%%%%%
%
\subsection{Experimental Quantum Simulations -\\
designing quantum many-body effects}
%
%Optically trapping and its prospects and limitations for Quantum
%Information Processing (QIP) or Quantum Metrology are not co
We sketch how the field of Analog Quantum Simulations (AQS) might be
enriched by optically trapped ions and atoms in optical lattices,
originally considered for the field of QIP\,\cite{Zoller2000}.
Optical lattices can be further stabilized and enhanced by optical
cavities in more dimensions while benefitting from the coordinated
choice of magic wavelengths\,\cite{Kaur2015} and state dependent
potentials\,\cite{Schneider2010}. For a more detailed and general
overview in the context of our proposal, please refer
to\,\cite{Schaetz2007,Schmitz2009,Schneider2012b,Bermudez2012,Clos2016,Cirac2012}
and references therein.
However, AQSs based on atoms and ions are already appealing in the
presence of rf-fields. The possibility to study dynamics from
Peierls-transitions\,\cite{Bissbort2013} to friction models with a
very high degree of control is
intriguing\,\cite{Gangloff2015,Bylinskii2015,Bylinskii2016}. One
option is to tailor non-harmonic potentials by a combination of
optical lattice potential and rf-trap confinement, as suggested
in\,\cite{Hauke2010}, here, to study quantum heat engines, as well
as motors and ratchets in such potentials,
while considering quantum tunneling.\\
Realizing the proposal of Richard Feynman, that is, to run AQSs of
mesoscopic size, is predicted to open new frontiers of science by
studying high-energy physics, cosmology, atomic physics, quantum
chemistry and even biology. With the envisioned size of only tens of
ions/atoms/spins, we could already perform useful AQS, beyond the
scope of classical
computation\,\cite{Doerk2010,Cirac2012,Eisert2015}.
In this context we propose to exploit the major advantage of an
array of ions compared to atoms in an optical lattice, that is, that
the effective interactions are long-range, since they are mediated
by the Coulomb force. Furthermore, we aim to circumvent at a stroke
the major disadvantage of the ionic-approach, its current lack of
scalability in size and dimension. In addition, new classes of AQS
will become accessible, by combining optically trapped ions and
atoms in common optical lattices. It might be possible, for example,
to study atoms in a completely occupied lattice, enriched by a small
density of ions, sharing electrons by tunnelling, causing highly
entangled states of the compound system featuring quantum dynamics
governed by the Bose-Hubbard Hamiltonian.
Further scaled AQS would not only provide new results that cannot be
predicted otherwise or classically simulated, but they would also
allow for the test of various models, required for substantially
deepening our understanding of complex quantum dynamics. Here, the
proposed scaling in optical lattices might feature a controllable
version of magnified lattice structures, providing clean control of
otherwise not accessible quantum effects in solids and could be used
to study problems in condensed matter physics, such as correlated
electrons or quantum magnetism, High-Tc superconductors, quantum
Hall ferromagnets, ferroelectrics etc. Their experimental simulation
would allow to observe and analyze quantum phase transitions.\\

This review presents complementary approaches identifying intriguing
and challenging problems related to optically trapping of ions as
well as early experiments. Currently the field is driven by rapid
development and active research. Therefore, the larger picture is in
no way complete and we envision the most exciting developments and,
most importantly, the related results providing deeper insight into
complex quantum behavior, in the (near) future.\\
%%%%%%%%%%%%%%%%%%%%%%%%%%%%%

We thank I.Cirac and D.\,Leibfried for fundamental contributions and
J.\,Dallibard, M.\,Drewsen, R.\,Moszynski, R.\,Cote,  G.\,Morigi,
V.\,Vuletic and P.\,Zoller for fruitful discussions and their
comments. This project has received funding from the European
Research Council (ERC) under the European Union's Horizon 2020
research and innovation program (grant agreement n$^\circ$ 648330).
We acknowledge support from the DFG within the GRK 2079/1 program
and DFG 91b support (INST 39/828-1 and 39/901-1 FUGG), a stipend by
the Studienstiftung des deutschen Volkes and from Marie Curie
Actions.
\bibliography{ref}
\end{document}